\documentclass[twocolumn,tighten]{aastex63}

\usepackage{times,natbib,graphicx,amsmath,multirow,xspace}
\usepackage{xcolor}
\usepackage{lineno}
\let\tablenum\relax
\usepackage{siunitx}
\usepackage{hyperref}

\newcommand{\nustar}{NuSTAR\xspace}

\newcommand{\nicer}{NICER\xspace}

\newcommand{\ms}{$M_{\odot}$\xspace}

\newcommand{\rin}{$R_{\rm in}$\xspace}
\newcommand{\rg}{$R_{g}$\xspace}
\newcommand{\risco}{$R_{\mathrm{ISCO}}$\xspace}

\newcommand{\relxillns}{{\sc relxillNS}\xspace}

\shorttitle{NICER-NuSTAR View of Serpens X-1}
\shortauthors{Hall et al.}

\begin{document}

\title{Simultaneous \nicer and \nustar Observations of the Neutron Star Low-mass X-ray Binary Serpens X-1}

\correspondingauthor{H. Hall}
\email{hk5269@wayne.edu}

\author[0009-0000-4409-7914]{H. Hall}
\affiliation{Department of Physics \& Astronomy, Wayne State University, 666 West Hancock Street, Detroit, MI 48201, USA}
\author[0000-0002-8961-939X]{R.~M.~Ludlam}
\affiliation{Department of Physics \& Astronomy, Wayne State University, 666 West Hancock Street, Detroit, MI 48201, USA}
\author[0000-0003-2869-7682]{J. M. Miller}
\affiliation{Department of Astronomy, University of Michigan, 1085 South University Ave, Ann Arbor, MI 48109-1107, USA}
\author[0000-0002-9378-4072]{A. C. Fabian}
\affiliation{Institute of Astronomy, University of Cambridge, Madingley Rd, Cambridge CB3 0HA, UK}
\author[0000-0001-5506-9855]{J. A. Tomsick}
\affiliation{Space Sciences Lab, University of California, Berkeley, 7 Gauss Way, Berkeley, CA, 94720, USA}
\author[0000-0001-7532-8359]{J. Coley}
\affiliation{Department of Physics and Astronomy, Howard University, Washington, DC 20059, USA CRESST and NASA Goddard Space Flight Center, Astrophysics Science Division, 8800 Greenbelt Road, Greenbelt, MD 20771, USA}
\author[0000-0003-3828-2448]{J. A. Garc\'ia}
\affiliation{Cahill Center for Astronomy and Astrophysics, California Institute of Technology, 1200 E. California Blvd, MC 290-17, Pasadena, CA, 91125, USA}
\author[0000-0003-0870-6465]{B. M. Coughenour}
\affiliation{Department of Physics, Utah Valley University, 800 W. University Parkway, MS 179, Orem, UT 84508, USA}

\begin{abstract}
We present the first contemporaneous \nicer and \nustar analysis of the low-mass X-ray binary Serpens X-1 obtained in June 2023, performing broadband X-ray spectral analysis modeling of the reprocessed emission with \relxillns from $0.4-30$ keV. We test various continuum and background estimation models to ensure that our results do not hinge on the choice of model used and found that the detection of reflection features is independent of the choice of both continuum and background model. The position of the inner accretion disk is consistent with the last stable circular orbit ($R_{\rm in} \leq 1.2$~\risco) and a low inclination of $i\leq 8.3 ^{\circ}$. 
Additionally, we investigate the presence of the low energy ($\sim$ 1 keV) Fe L complex in the data from \nicer and the Reflection Grating Spectrometer (RGS) on XMM-Newton that was previously reported in the literature. We find that the line is at most a 2\% feature relative to the reprocessed continuum and are unable to claim a definitive detection for the current dataset. However, we discuss plausible conditions and systems that would increase the likelihood of detecting this feature in the future.

\end{abstract}

\keywords{accretion, accretion disks --- stars: neutron --- stars: individual (Serpens X-1) --- X-rays: binaries}

\section{Introduction} \label{sec:intro}
Neutron star (NS) low-mass X-ray binaries (LMXBs) are systems comprised of a NS and a companion star that is $\lesssim$ 1 \(M_\odot\). The NS accretes matter via a disk formed from Roche-Lobe overflow of the companion. The accretion disk can be illuminated by hard X-rays originating from either a hot electron corona \citep{1991SvAL...17..409S} or the surface of the NS or the boundary layer (BL) \citep{Popham_2001}. The hard X-rays are reprocessed by and re-emitted from the accretion disk in the form of a series of atomic features and a Compton backscattering hump that appear superimposed on a reprocessed continuum. This is widely referred to as the ``reflection" spectrum of the system. It has been shown that this reflection spectrum can give valuable insight into properties about the NS and accretion disk (see \citealt{ludlam24} and references therein for a recent review). 

Within the LMXB classification, sources can be further separated into ``atoll" and ``Z" sources, based on the shape drawn out in their hardness-intensity diagrams (HIDs) and color-color diagrams (CCDs). ``Z" sources are generally more luminous, often around or even exceeding the Eddington limit ($L_{\rm Edd}$) and may show three distinct branches (Normal, Horizontal, and Flaring) in their HID and CCD \citep{1989A&A...225...79H}.
``Atoll" sources typically output $\sim$ 0.01-0.5 $L_{\rm Edd}$ and can be either found in the so-called soft ``banana" or hard ``island" states \citep{Homan_2010}. 
Atoll sources often show a softer spectrum at high luminosities and a harder spectrum at low luminosities. However, within the soft (banana) state, the hardness ratio remains relatively constant across a variety of luminosities \citep{Church_2014}. 

Serpens X-1 (or Ser X-1) is a bright, persistently accreting atoll source estimated to be at a distance of 7.7 $\pm$ 0.9 kpc that has only been observed in a  soft state corresponding to the ``banana" branch \citep{Galloway_2008}. The source has exhibited many Type-I X-ray bursts, including evidence of a `superburst' \citep{Galloway_2008,2002_Cornelisse}. Serpens X-1 has been reported to have an inner-disk radius ranging from $\sim$ 7 -- 26 \(R_g\)\footnote{\rg = $GM/c^2$} \citep{2010_Cackett_et_al,Miller_2013,Matranga_2017,chiang2016evolution,Chiang_2016_FeK}. The lower inferred radii imply that the inner accretion disk radius is truncated primarily due to a boundary layer as opposed to a strong magnetic field \citep{chiang2016evolution}. It has been observed with every major X-ray mission (see \citealt{Mondal_2020, Ludlam_2018} for an extended list). The binary system has been shown to have a low inclination of (i $\le 10^{\circ}$, \citealt{Cornelisse_2013, Miller_2013, Ludlam_2018}), however other studies have given a range of $25^{\circ} <$ i $< 50^{\circ}$ (\citealt{Cackett_2008, 2010_Cackett_et_al, Matranga_2017, chiang2016evolution}).

Among these numerous observations Serpens X-1 has shown persistent evidence of emission lines, most notably the prominent Fe K$\alpha$ reflection feature, between 6.4--6.97 keV. Serpens X-1 was the first NS LMXB to show confirmed evidence of a relativistically broadened reflection feature, expanding the field of reflection studies, which up to that point had been utilized solely on black hole systems \citep{Bhattacharyya_2007, Cackett_2008}. In addition, there has been evidence of a lower energy feature present in the reflection spectrum, namely the Fe L complex near 1 keV \citep{Ludlam_2018}. 
The Fe L feature has been observed from accreting supermassive black holes when iron abundance in the disk is high \citep{2009Natur.459..540F}, but can also be generated from high disk density effects \citep{Garc_a_2022}.
It is worth noting that the feature we refer to as Fe L is more than just a single broadened emission line but rather a mix of lower Z-elements in addition to the Fe L-shell transition \citep{Ludlam_2018}. 
It was shown that this emission feature in Ser X-1 is also relativistically broadened, similar to the Fe K line \citep{Ludlam_2018}. A 1 keV line has also been noted in several other NS LMXBs \citep{2010_Cackett_et_al,Iaria_2009,Di_Salvo_2001,Sidoli_2001,Boirin_2004,M_ck_2013,Ng_2010,2009Natur.459..540F}. Recently, a low-energy line has been seen in Cygnus X-2 as reported by \cite{2022_Ludlam} yet the most likely explanation of its existence is from either photoionization or collisionally ionized material from further out in the disk. The origin of this lower energy emission feature is still not yet concretely known for X-ray binaries, but an upcoming paper seeks to provide a coherent picture \citep{2024_Chakraborty}. 

The combination of \nustar and \nicer observations is ideal for studies of reprocessed emission due to the overlapping coverage of the Fe K region (5--7 keV), as well as the complementary energy bands that allow for the analysis of both the hard and soft X-ray emission. We make particular note that both instruments are free from pile-up effects, an issue that has plagued CCD detectors in the past. Event pile-up is known to cause flux loss and shift the overall shape of the spectrum, resulting in a distortion of potential spectral features \citep{2001_Davis,2015_Jethwa,2010_Miller,Ng_2010}. The event thresholds and readout times on \nicer and \nustar prevent these effects, provided the source is not exceptionally luminous. In the present work we analyze simultaneous \nustar and \nicer data of Serpens X-1. We seek to investigate the presence of the low energy feature as reported by \citet{Ludlam_2018} in addition to further constraining the possible geometry of the source (inclination, inner disk radius, etc.). We place this in the context of previous detection in archival \nicer data as well as the XMM-Newtown Reflection Grating Spectrometer (RGS) data used in \citet{Ludlam_2018}. We also test the impact that available \nicer background estimation models may have in modeling the source data. This study represents the first broad passband (0.4-30 keV) analysis of Serpens X-1 devoid of the aforementioned pileup effects leading to a more robust and accurate treatment of the entire spectrum.

\section{Observations and Data Reduction} \label{sec:data}
\subsection{NICER}
\nicer observed Serpens X-1 three times over a span of three days beginning on 2023 June 02 (ObsIDs 6648010101-103) for a cumulative total of 23 ks of exposure time. The data were reduced using the standard {\sc nicerl2} for calibration and filtering with CALDB version xti20221001. Good time intervals (GTIs) were created using {\sc nimaketime} with the COR\_SAX$\geq$4, in an effort to remove intervals with high particle background radiation, as well as KP$\le$5, an indicator of the strength of the geomagnetic field. These GTIs were applied to the data through {\sc niextract-events} selecting events with PI channel between 35 and 1200 (0.35--12.0 keV).  
The event files were then read into \textsc{xselect}, creating light curves in super-soft (0.5--1.1 keV), soft (1.1--2.0 keV), intermediate (2.0--3.8 keV), and hard (3.8--6.8 keV) energy bands, as defined in \citet{2018_Bilt_et_al}. No Type-I X-ray bursts were seen in the 1-second light curves so no additional filtering was needed.

\begin{figure}[t]
    \centering
    \includegraphics[width=\linewidth]{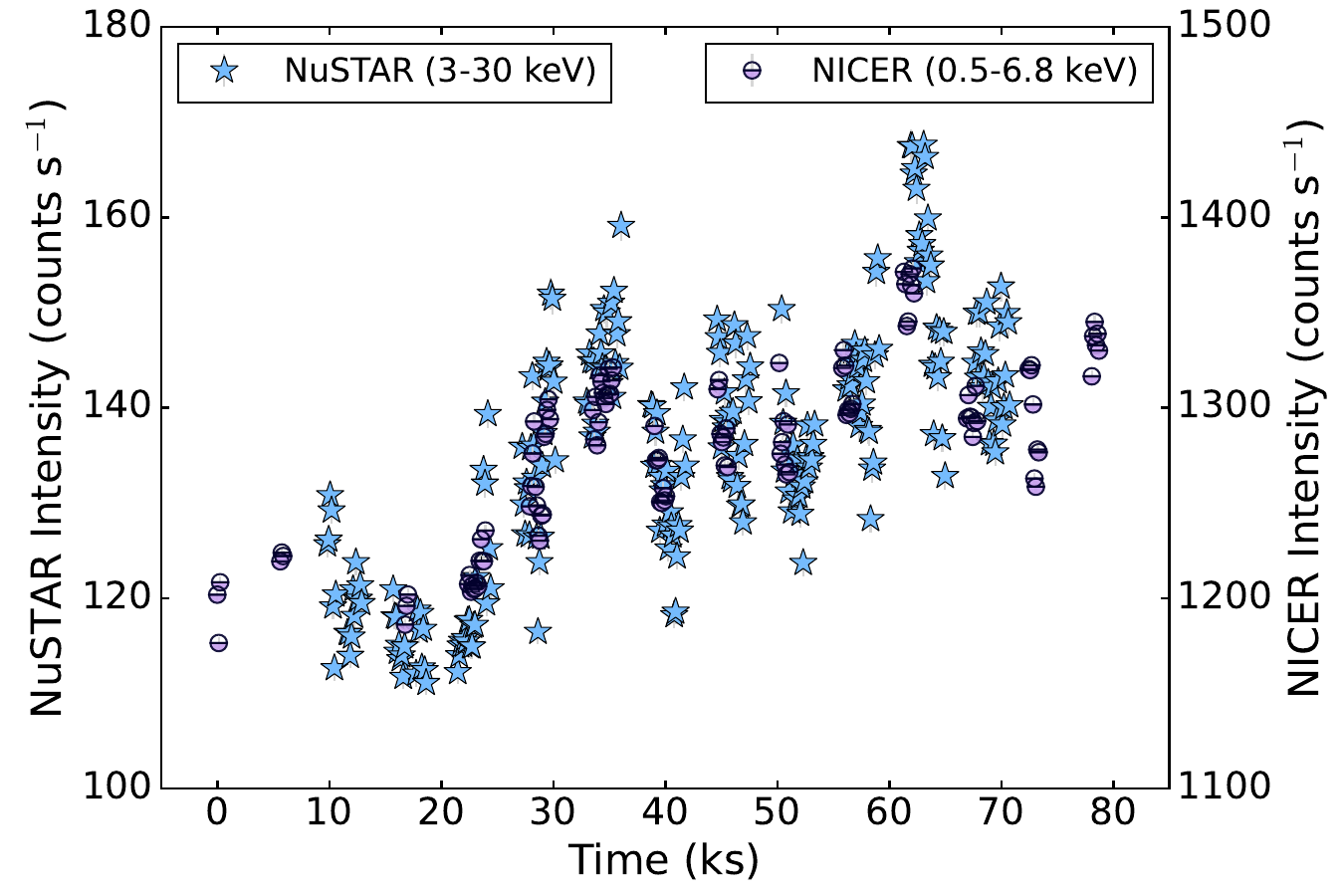}
    \caption{The lightcurve of Serpens X-1 of the joint observations binned at with 128 second bins. The lavender circles represent the \nicer observations and the blue stars represent data taken by \nustar. Time in ks is plotted on the x-axis while NICER and NuSTAR intensity are plotted on the y-axis.}
    \label{2023 NINU LC}
\end{figure}

Background spectra, source spectra, and response files were created using the {\sc nicerl3-spect} task which adds systematic errors from the CALDB. We selected the optimally binning option with a minimum number of 25 counts per bin to employ $\chi^2$ statistics. See \citet{Kaastra_2016} for more on the optimal binning scheme. The default background model used by {\sc nicerl3-spect} is the `SCORPEON' model but it also allows for the creation of background using the `3C50' \citep{Remillard_2022} and `Space Weather' models. Background spectra files were created using SCORPEON and 3C50 models for comparison of their effect on the resulting spectra. One of the primary motivations behind this comparison was the fact that the SCORPEON background was designed as a model that fits alongside the continuum model as opposed to a static background file. We aim to see what effect, if any, this will have on the resulting residuals.

\subsection{NuSTAR}
\emph{\nustar} observed Ser X-1 for 24.7 ks on 2023 June 03 (ObsID 30901004002). We ran {\sc nupipeline} from {\sc nustardas} v2.1.2 with CALDB 20230530 to reprocess the data with statusexpr = ``(STATUS==b0000xxx00xxxx000)\&\&(SHIELD==0)" since the source exceeds 100 counts s$^{-1}$. We created \SI{100}{\arcsecond} circular extraction regions in {\sc ds9} around the source to produce source spectra for both FPMA and FPMB detectors. Another circular \SI{100}{\arcsecond} region was used away from the source for the purpose of background subtraction. Light curves and spectra were generated via {\sc nuproducts}. Upon inspection of the 1-second light curves, three Type-1 X-ray bursts had occurred during the observation. These were filtered out using good time intervals (GTIs) that started a few seconds prior to the burst and continued until the emission returned to the persistent emission level ($\sim60$ seconds total). These GTIs were given to {\sc nuproducts} to extract light curves and spectra of the persistent emission with the bursts removed. These are the files we use as the analysis of the burst emission is outside the scope of the paper. 

The persistent spectra were then optimally binned with a minimum count of 25 using {\sc ftgrouppha}.  

\subsection{XMM-Newton/RGS}
We use the three observations from XMM-Newton of Ser X-1 in March 2004 (ObsIDs 0084020401, 0084020501, 0084020601) with a total combined exposure of $\sim$ 65.7 ks. These observations were previously reported in \citet{Bhattacharyya_2007, 2010_Cackett_et_al, Matranga_2017,Ludlam_2018}. In the effort to verify and match the findings of \citet{Ludlam_2018}, we use the RGS data for its high resolution data in the lower energies around the 1 keV feature. To maintain consistency, the data were reduced according to the procedure described in \citet{Ludlam_2018}, using the {\sc rgsproc} command in SAS v21.0. The first order RGS1 and RGS2 data were combined using {\sc rgscombine} for each observation. The resulting spectra were optimally binned with a minimum count of 25 using {\sc ftgrouppha}. Additionally, we utilized the ``analysis date" option when creating the ``current calibration files" to investigate any possible differences between the current up-to-date calibration of the instrument and that used in \citet{Ludlam_2018}\footnote{The analysis date corresponding to the current paper was 27 August, 2024. For the previous study we used the date of draft publishing on arxiv.com (2 May, 2018)}. 

\begin{figure}[t]
    \centering
    \includegraphics[width=\linewidth, trim=10 20 1 5, clip]{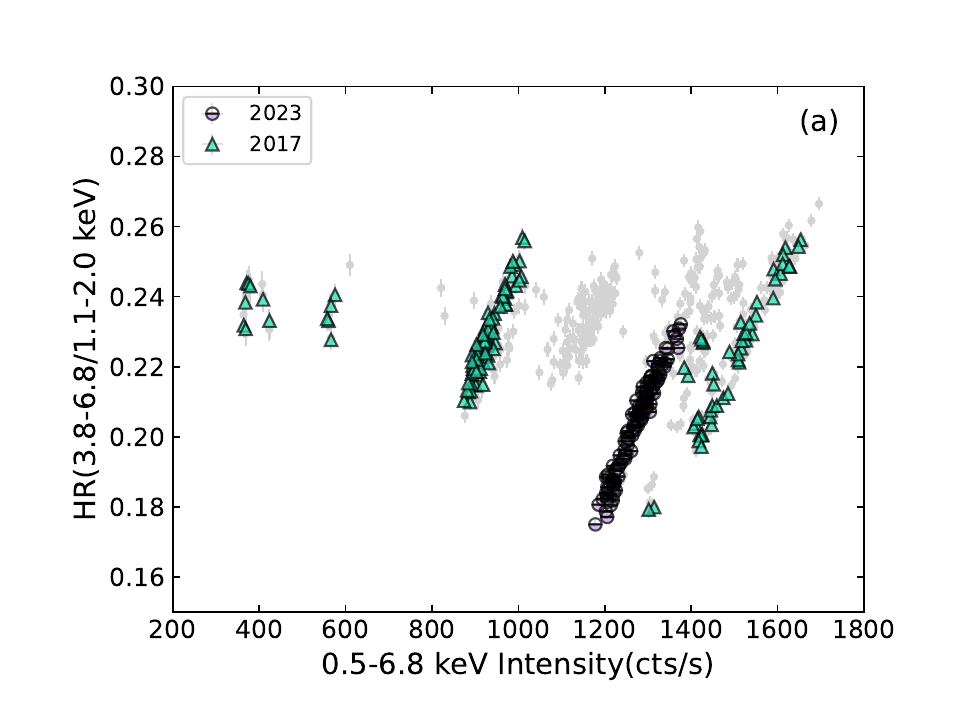}
    \includegraphics[width=\linewidth, trim=10 0 1 20, clip]{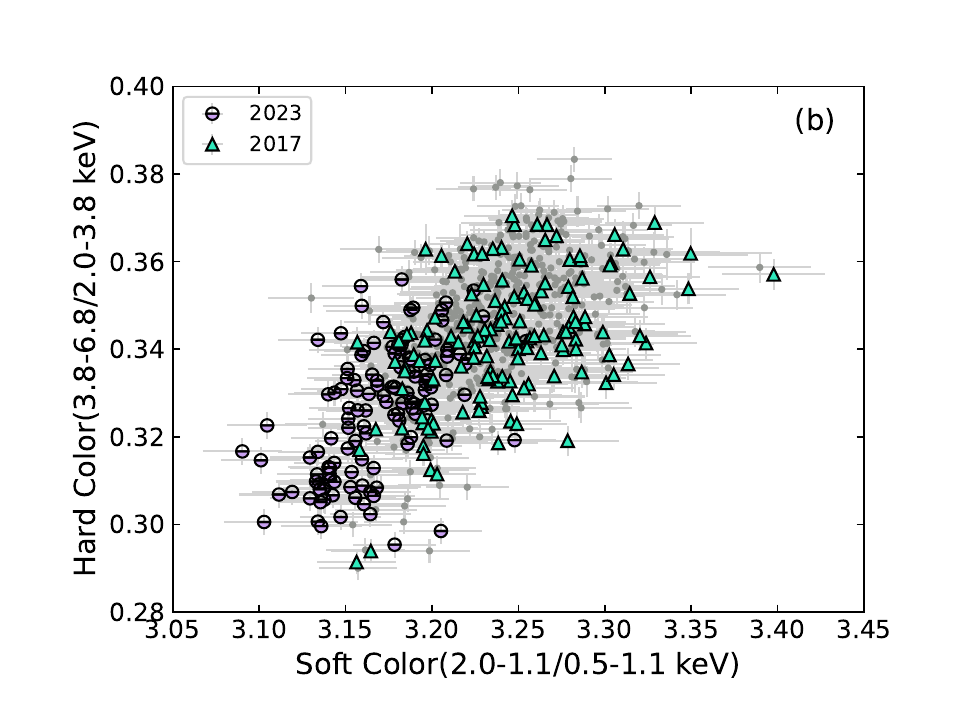}
    \caption{Panel (a): NICER HID for Serpens X-1. The 2017 data (aquamarine triangles) can be seen spanning a wide range of intensities while the 2023 observation (lavender circles) occupies an average range in between. We find the HR to be comparable. Panel~(b): NICER color-color Diagram for Serpens X-1. The 2023 observation being reported on here is seen as lavender circles, 2017 data set reported in \citet{Ludlam_2018} are represented by aquamarine triangles, and all other archival data appear as grey circles. 
    All other archival data is marked as grey circles.}
    \label{NICER HID}
\end{figure}

\section{Analysis and Results} \label{sec:results}
\subsection{Light Curve Analysis}
Figure \ref{2023 NINU LC} shows the \nicer and \nustar light curves during the contemporaneous observation. Figure \ref{NICER HID} panel (a) shows the hardness-intensity diagram (HID) for all archival \nicer data on Serpens X-1. The data span a wide range of intensity, however, the hardness ratio (HR) remains relatively stable. We can conclude that the source was not in the hard island state during the 2023 observations. It is worth noting that Serpens X-1 does appear to be in a slightly softer state on average in comparison to the 2017 data reported in \citet{Ludlam_2018}. 

\begin{figure}[t]
    \centering
    \includegraphics[width=\linewidth, trim=10 10 1 5, clip]{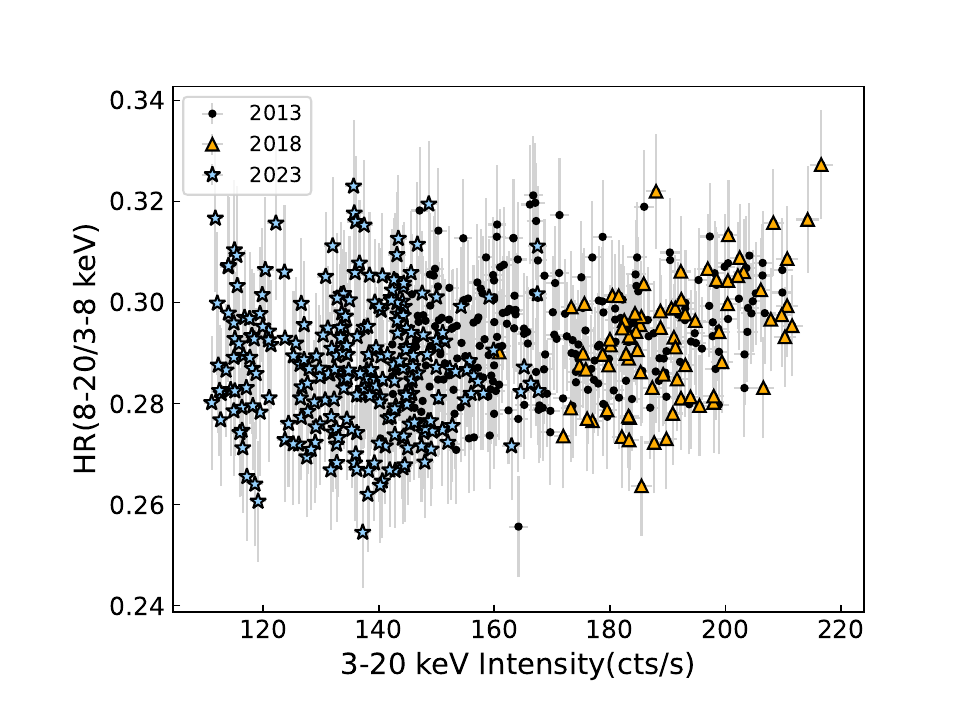}
    \caption{NuSTAR HID, where 2013 data reported in \cite{Miller_2013} appear as black circles, 2018 data reported in \cite{Mondal_2020} appear as orange-yellow triangles, present observations appear as blue stars.}
    \label{NUSTAR HID}
\end{figure}

The CCD, Figure \ref{NICER HID} panel (b), shows a similar overlapping pattern; however the 2023 data shows lower average color ratios for both hard and soft color. But together they do not stray far from the overall average, taking into account all archival data. Figure \ref{NUSTAR HID} shows the HID for NuSTAR with the hardness ratios being comparable and the only noticeable difference being the overall intensity in the 2023 data.  It is well within the average of all total observations despite being on the lower end, and the 2018 data that was reported in \cite{Mondal_2020} being on the more luminous end.

\subsection{\nustar and \nicer Spectral Modeling}
We consider the \nicer data from 0.4--10 keV and the \nustar data in the energy range of 3--30 keV as the spectra became background dominated beyond this upper limit.
We account for the calibration differences in the respective missions using {\sc crabcor} in {\sc xspec}. Not a native model to {\sc xspec}, {\sc crabcor} is a multiplicative model component comprised of a normalization constant and an $E^{\Delta\Gamma}$ term originally introduced in \citet{2010_Steiner_et_al}. The normalization constant is fixed to 1 for the FPMA spectrum of \nustar and allowed to vary for the FPMB and \nicer spectra to account for flux calibration differences. The $\Delta\Gamma$ term is set to zero for both \nustar spectra, and allowed to vary for \nicer to account for inherent spectral slope differences between missions.
We account for the neutral column density along the line of sight with {\sc tbfeo}. Within {\sc xspec}, we set abundances to {\sc wilm} \citep{2000_Wilms} and used {\sc vern} cross sections \citep{1996_Verner}. Additionally, we used two \textsc{edge} components at $\sim0.445$ keV and between 0.8 -- 0.9 keV, to account for low energy features in the \nicer spectra similar to \citet{Ludlam_2021, Moutard_2023} that are likely astrophysical in origin. Errors are reported at the 90$\%$ confidence level from a Markov Chain Monte Carlo (MCMC) with 200 walkers, a burn length of $2 \times10^4$, and a chain length of $1\times10^5$. 

\subsubsection{Continuum Modeling}
There are two models commonly used in literature to fit the spectrum of Serpens X-1. Model 1 is comprised of a multicolor disk blackbody ({\sc diskbb}), a single-temperature blackbody ({\sc bbody}), and power law ({\sc powerlaw}) component \citep{Cackett_2008,Miller_2013,Chiang_2016_FeK,Ludlam_2018}. Model 2 uses {\sc diskbb}, and a thermal Comptonization component arising from a single-temperature blackbody ({\sc nthcomp}) \citep{Bhattacharyya_2007,chiang2016evolution,Mondal_2020}. 

\begin{table}[!t]
\begin{center}
\caption{NICER/NuSTAR Continuum Model Fits}
\label{tab:cont_model_fit}
\begin{tabular}{llcc}

\hline
\textbf{Component}                  & Parameter         &     Model 1   & Model 2\\
\hline
\textsc{crabcor}    & $C_{FPMB}$ $(10^{-1})$ &$9.89\pm0.01$  & $9.90\pm0.01$   \\
& $C_{NICER}$ $(10^{-1})$ &$8.11\pm0.01$ & $8.2\pm0.1$   \\
& $\Delta\Gamma$ $(10^{-2})$ &$-8.6\pm0.8$   & $-7.6\pm1.0$ \\
\textsc{tbfeo}       & $N_{H}(10^{21}\ \rm cm^{-2})$ &$6.9^{+0.3}_{-0.1}$ &   $5.3^{+0.2}_{-0.3}$      \\
&$A_{O}$          &$1.5\pm0.1$      & $1.5\pm0.1$  \\
&$A_{Fe}$         &$2.9\pm0.2$      &  $2.2\pm0.3$  \\
\textsc{edge}            &E ($10^{-1}$ keV) & $8.9\pm0.2$ &  $8.6\pm0.2$\\
&$\tau_{max} (10^{-1})$            & $2.6\pm0.3$   &  $1.4\pm0.3$ \\
&E ($10^{-1}$ keV) & $4.43_{\emph{f}}$ & $4.43_{\emph{f}}$\\
&$\tau_{max}$  &$2.12^{+0.13}_{-0.27}$ & $1.28^{+0.26}_{-0.23}$\\
\textsc{diskbb} & $kT$ (keV) &$1.76\pm0.02$ & $1.98\pm0.03$ \\
&norm       &$25\pm 1$   &  $10\pm1$  \\
\textsc{bbody} &$kT$ (keV)          & $2.46^{+0.02}_{-0.03}$ & ... \\
&norm$(10^{-2})$       & $1.5\pm0.1$ & ... \\
\textsc{powerlaw} & $\Gamma$         & $3.4^{+0.04}_{-0.05}$  & ... \\
&norm            & $2.4\pm0.2$     & ...   \\

\textsc{nthcomp} & $\Gamma$ & ...&   $2.07\pm0.03$  \\
&$kT_e$ (keV) & ...& $3.1\pm0.1$  \\
& $kT_{bb}$ (keV) &...  &  $0.11\pm0.01$  \\
& norm  & ... &   $1.3\pm0.1$  \\
\hline
$\chi^2$ (dof)             & &2026.95 (398) & 1957.46 (398)

\end{tabular}

\medskip 
\end{center}
\end{table}

Table 1 reports the parameter values for Model 1. We find a hydrogen column density $(N_H)$ of $6.9 \times 10^{21}$ cm$^{-2}$, which is higher than the value given by \citet{1990DickeyLockman} of $4 \times 10^{21}$ cm$^{-2}$, however it is well within previously reported values by \citet{Ludlam_2018,chiang2016evolution,Matranga_2017}. It has also been noted in \citet{Corrales_et_al_2016} that measured $N_H$ values from X-ray absorption will be at minimum 25\% higher than those found by \citet{1990DickeyLockman}. Letting this parameter fit freely allows for a more realistic determination of the other model parameters and as a result, our fit parameters generally align with other reported values that allow the column density to vary. We find absorption abundances of oxygen and iron in units of Solar abundances of 1.3 and 2.9, respectively. We report a disk temperature $1.76\pm0.02$ keV with a normalization of $25\pm1$ km$^2$/(D/10 kpc)$^2$ cos(i). 
The power law component of our model has a soft photon index of $\Gamma = 3.4_{-0.05}^{+0.04}$, which agrees with the previously found values \citep{Miller_2013,chiang2016evolution, Mondal_2020}.
We find an unabsorbed flux value of $6.79 \times 10^{-9}$ erg cm$^{-2}$ s$^{-1}$ in the 0.4--30 keV band. Assuming a distance of 7.7$\pm$0.9 kpc and a canonical NS mass of 1.4 \(M_\odot\), we find an Eddington ratio L/L$_{Edd}$ of 0.27\footnote{L$_{Edd}$ $= 1.764\times10^{38}$ erg s$^{-1}$. Assuming canonical NS mass of 1.4 \(M_\odot\) and pure ionized hydrogen.}, which as mentioned in \S 1, is consistent with values for atoll sources.

\begin{figure}
    \centering 
    \includegraphics[width=\linewidth, trim= 0 3 0 0, clip]{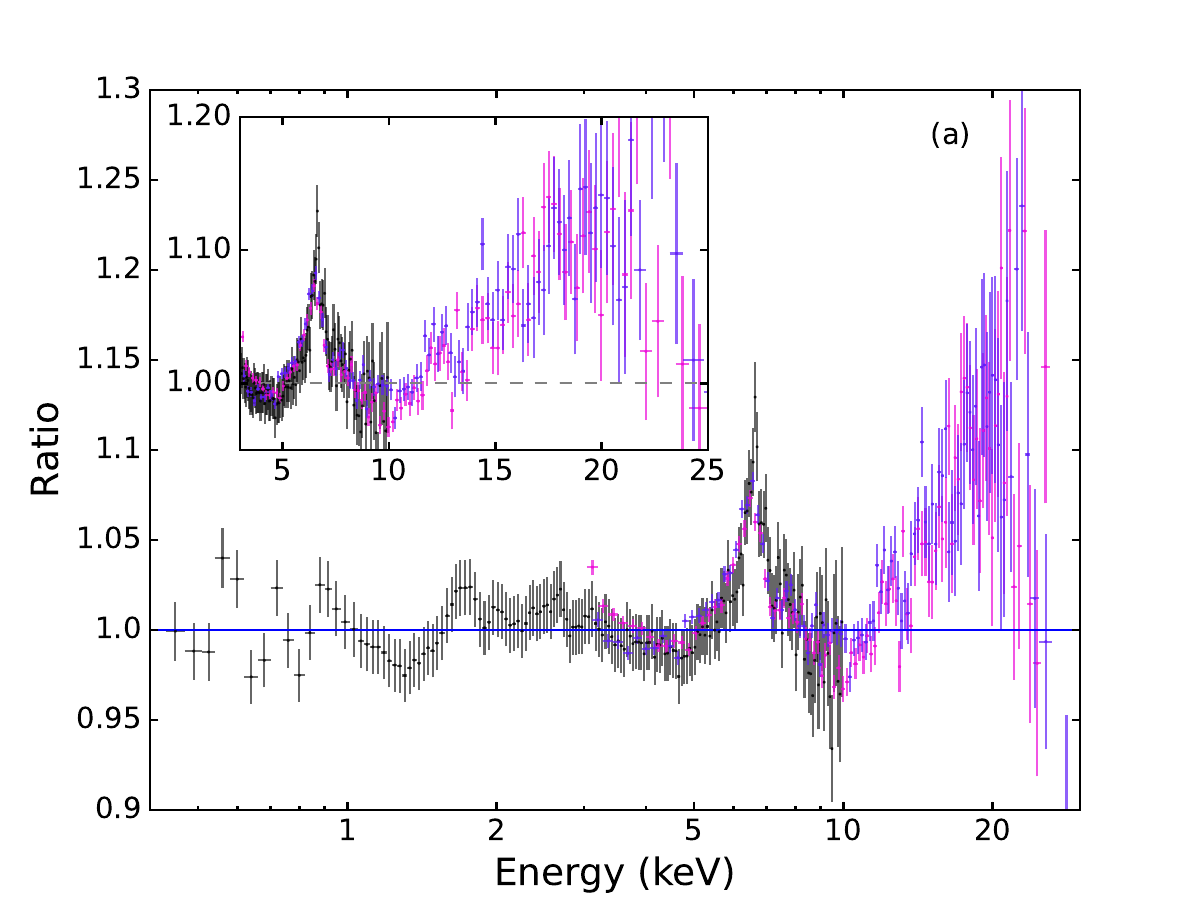}
    \includegraphics[width=\linewidth, trim= 0 0 0 7, clip]{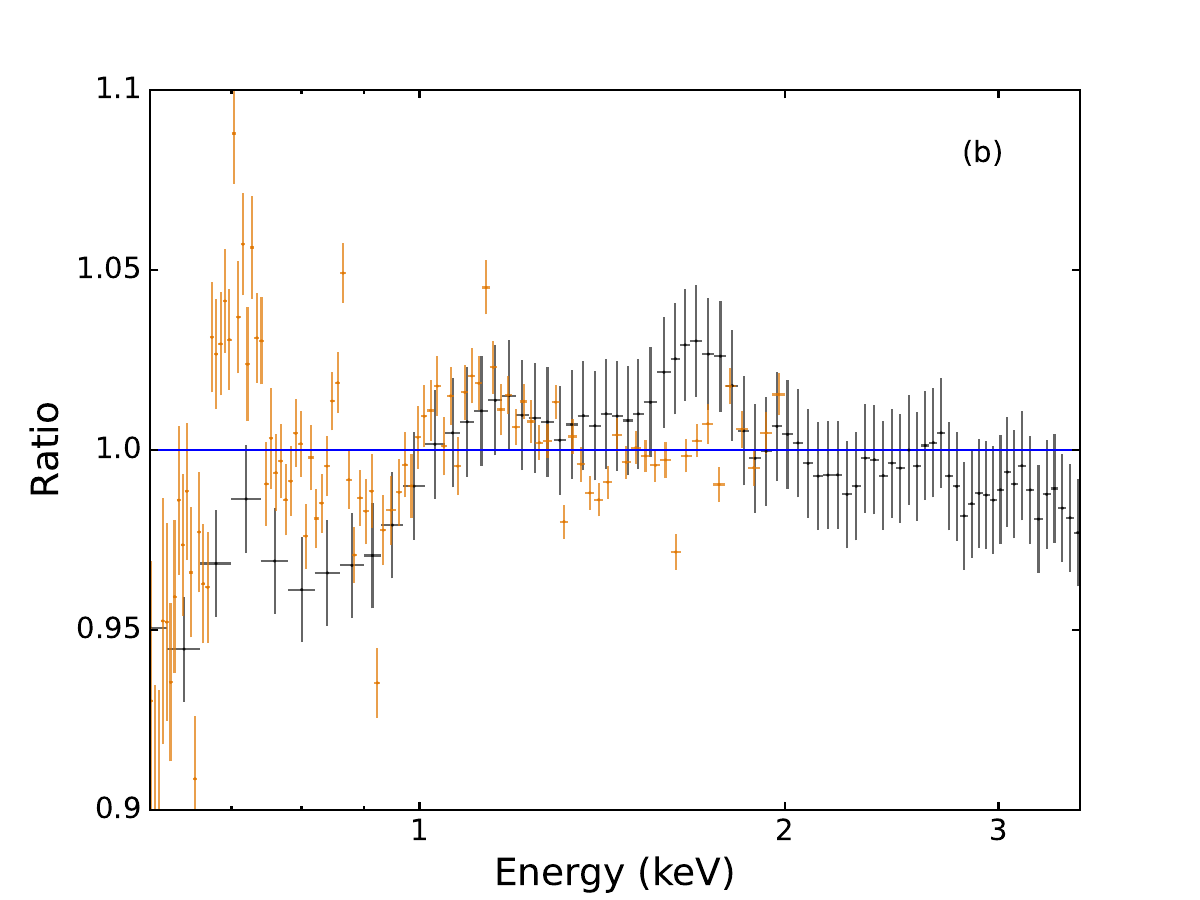}
    \caption{Ratio of the continuum fit for the contemporaneous 2023 \nicer (black), \nustar (FPMA:fuschia, FPMA:blue) and 2004 RGS (orange) observations of Serpens X-1. Panel (a) shows the NICER and NuSTAR, while panel (b) shows NICER and RGS from 0.5-3.5 keV to highlight the Fe-L complex. The Fe K$\alpha$ line and the Compton hump $\geq$ 10 keV are prominently shown in panel (a). The additional window in (a) displays the same data on a linear scale from 3 to 25 keV. Data has been rebinned for visual clarity.}
    \label{2023 Cont}
\end{figure}

Model 2 provided an improved fit ($\chi^2/d.o.f = 1957/398$) over Model 1. The model is able to more accurately describe the lower energy emission ($\le$8 keV) but provides a poorer description of the data at higher energies. Model 2 gives a hydrogen column density $(N_H)$ of $5.3 \times 10^{21}$ cm$^{-2}$, closer to but still higher than the value from \citet{1990DickeyLockman}. This model gives absorption abundances of oxygen and iron in units of Solar abundance of 1.5 and 2.2 respectively. Model 2 favors a slightly hotter disk with a disk temperature of $1.98\pm0.01$ keV and normalization of $10\pm1$~km$^2$/(D/10 kpc)$^2$ cos(i). We find a significantly harder photon index of  $\Gamma = 2.1$ from {\sc nthcomp} and the best fit gives an electron temperature $kT_e=3.1\pm0.1$ keV and a seed photon temperature $kT_{bb}= 0.11\pm0.01$ keV with a normalization of $1.3\pm0.1$. Given the lack of self-consistent reflection models for Comptonized continua from NSs (see \citealt{ludlam24} for a full discussion), we proceed with the Model 1 continuum description when modeling the reprocessed emission. Figure \ref{2023 Cont}(a) shows the ratio of the \nicer and \nustar data to Model 1 in the full 0.4--30 keV energy band.

\begin{figure}[t!]
    \centering
    \includegraphics[width=\linewidth, trim= 0 0 0 0, clip]{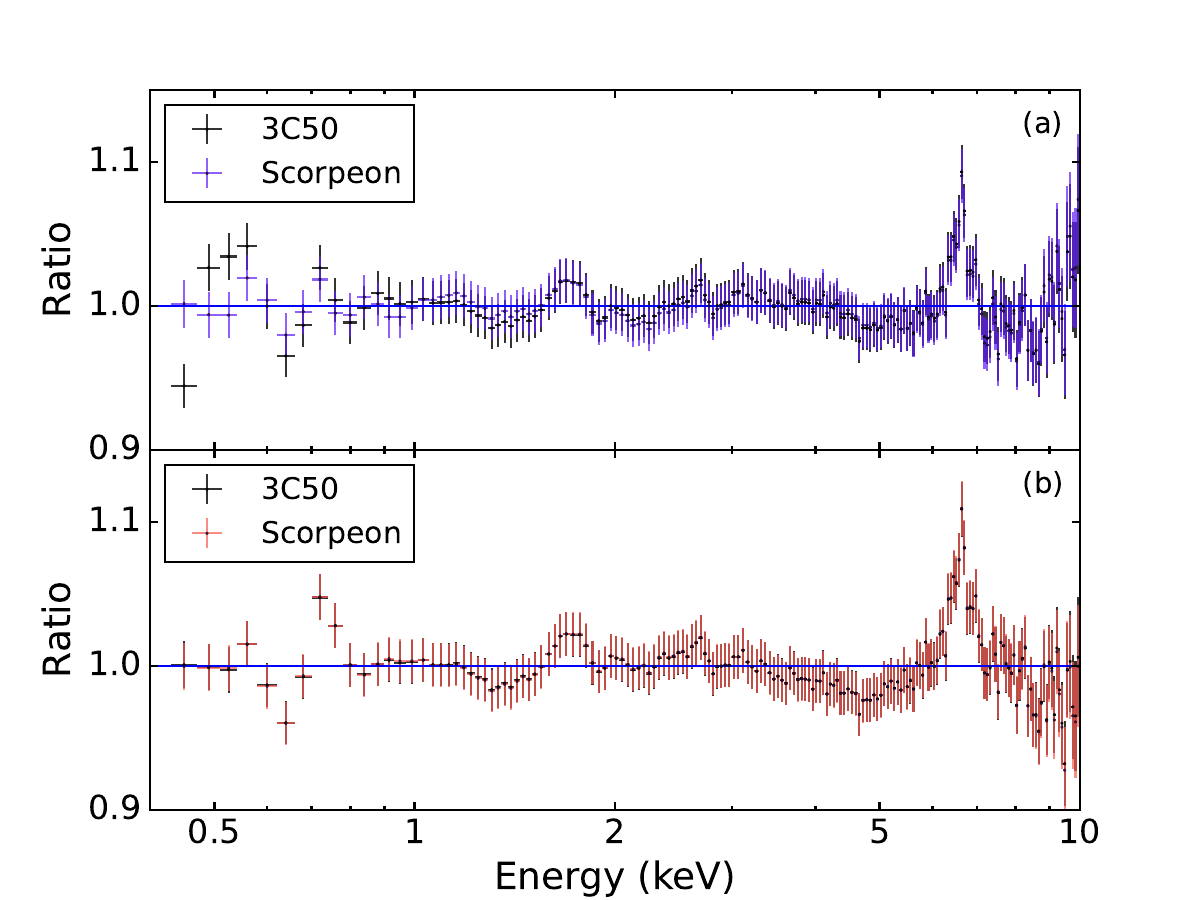}
    \caption{Ratio comparison of June 2023 NICER Observations. Panel (a) shows Model 1 as described in \S 3.2 while panel (b) shows the resulting ratio from Model 2. The 3C50 is represented by the black points and error bars in both plots, while SCORPEON is seen here in blue in panel (a) and coral in panel (b).}
    \label{fig 5}
\end{figure}

As mentioned in \S 2, two different background models were generated for the \nicer dataset to explore the potential effects on ``reflection" feature detection. In order to ensure that these features are not background or continuum model dependent, we conduct the continuum modeling with both background models. The previous values reported were using the `3C50' model.
A comparison of the ratio of the data to continuum model can be seen in Figure \ref{fig 5} for both `3C50' and the piece-wise `SCORPEON' background model. As can clearly be seen in the plot, the choice of background estimation appears to have very little effect on fit residuals. Thus, the choice of background-model does not affect the detection or classification of spectral features. Consequently, we chose to proceed with the `3C50' model for background files in this work. 

We highlight that there appear no notable differences between either background model specifically in the areas of interest near 1.1 keV and 6.4 keV, the regions where we see the Fe L and Fe K lines, respectively. We further note that while there is a prominent Fe K line, there is no apparent emission line near 1 keV for either continuum model regardless of background model used. We note that in all 2023 datasets a roughly 2.5\% feature can be seen at 1.74 keV.  
After consultation with the \nicer helpdesk, we've confirmed that it is the known Si edge of the instrument and therefore discount it as an astrophysical feature from the source itself.

\begin{figure}[t]
    \centering
    \includegraphics[width=\linewidth]{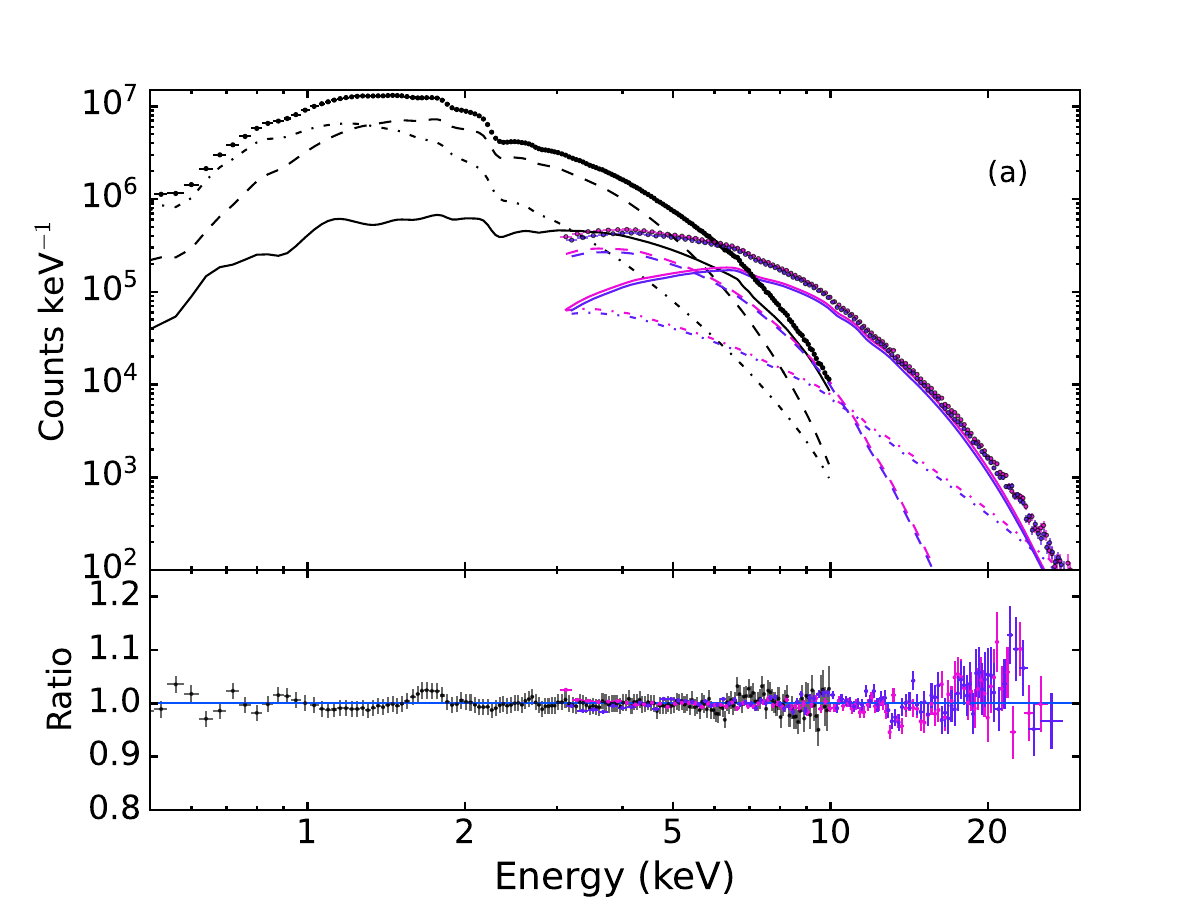}
    \includegraphics[width=\linewidth]{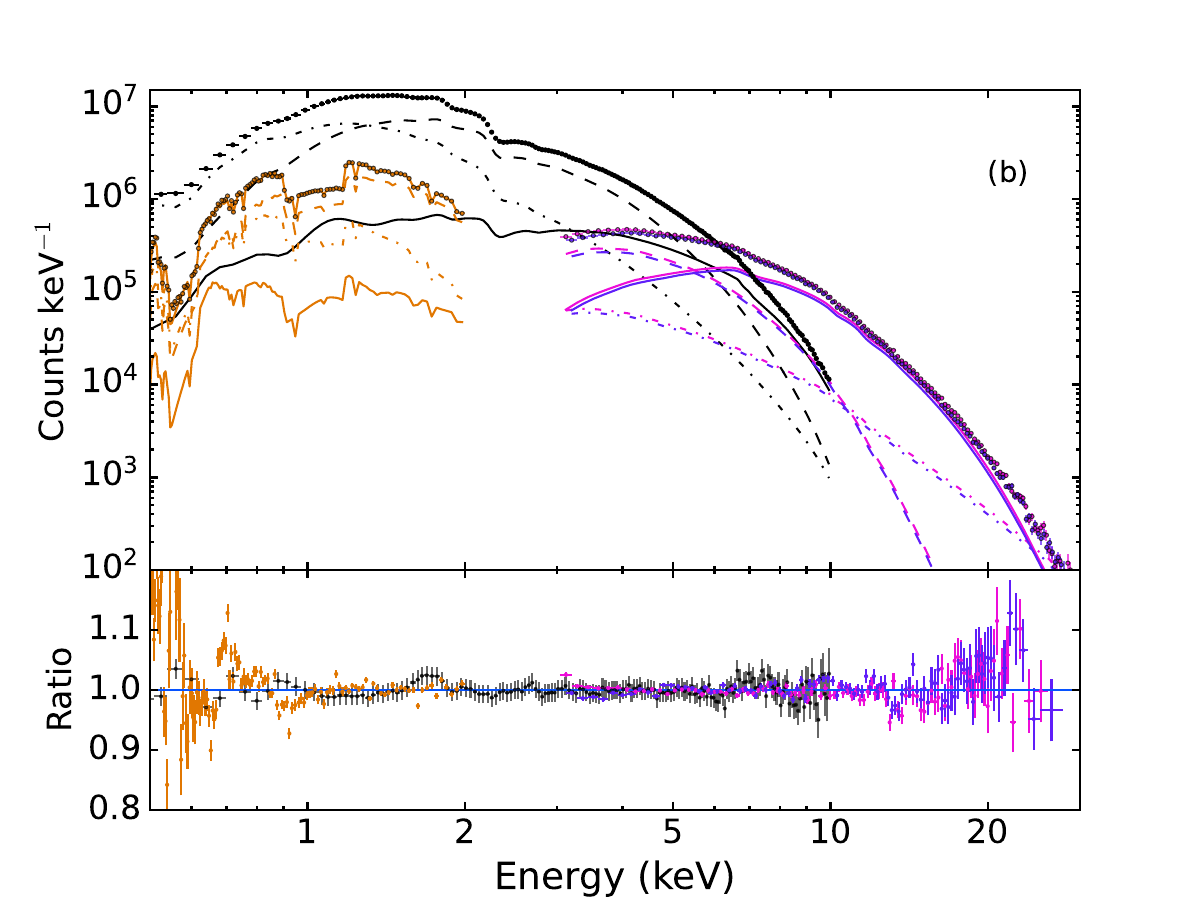}
    \caption{Panel (a) shows the spectrum and individual model components for \nicer (black) and \nustar (FPMA: fuschia, FPMB:blue) from 0.4-30 keV. \relxillns is seen here as the solid line, the disk blackbody is represented by the dashed lines, and the power law is seen here as the dot-dashed line. The lower panel shows the ratio of the data to model fit in Table 2 for the joint \nicer and \nustar observations of Serpens X-1. Panel (b) shows the addition of RGS (orange) to the reflection fit.}
    \label{refl_allcomp}
\end{figure}

\begin{figure}
    \centering
    \includegraphics[width=\linewidth]{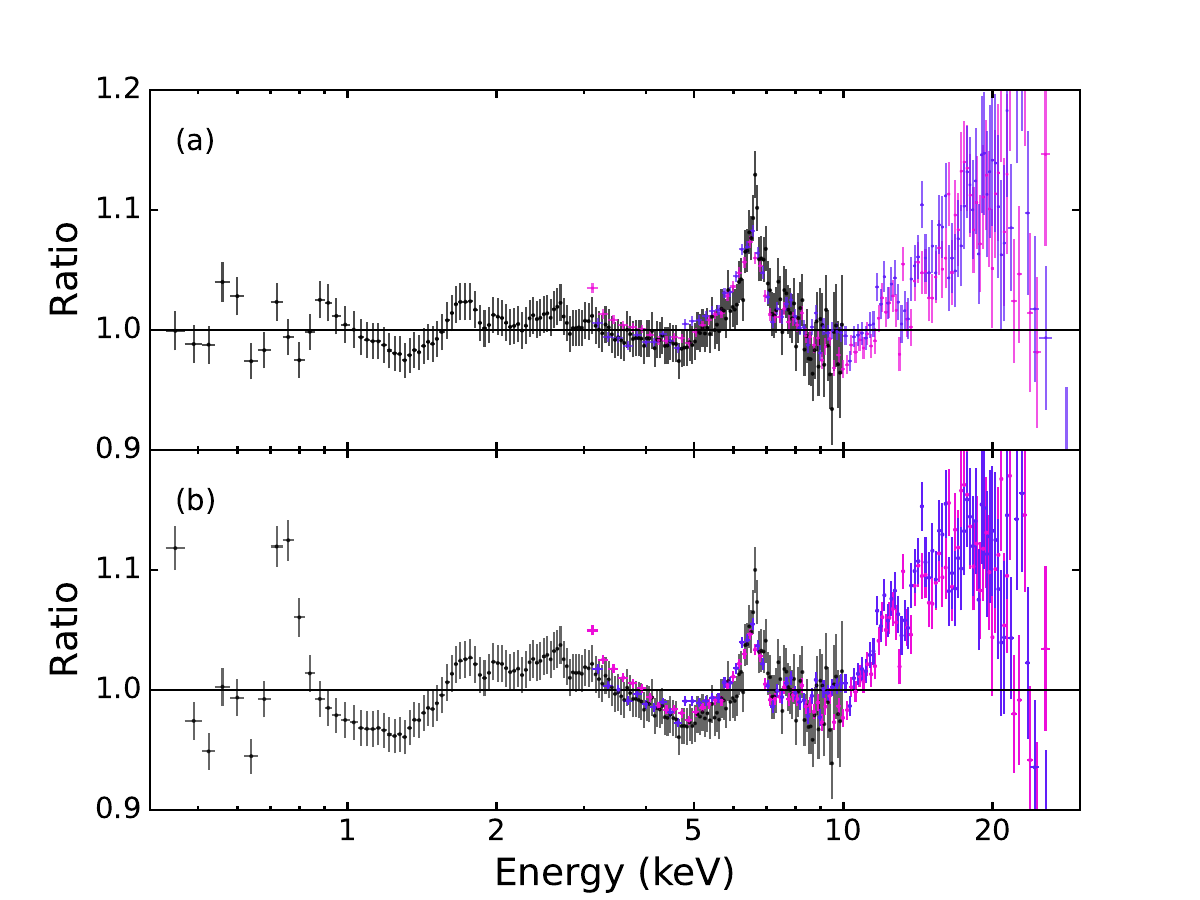}
    \caption{Ratio plots showing the comparison of best fits when including and excluding \textsc{edge} components in the continuum model. Panel (a) shows the resulting fit with edges included and panel (b) shows when the edges are removed. }
    \label{Edge Comparison}
\end{figure}

\begin{table}[h!t]
\begin{center}
\caption{NICER/NuSTAR Reflection Model Fit}
\label{tab:model_fit}
\begin{tabular}{llc}

\hline
Model                   & Parameter         & \\
\hline
\textsc{crabcor}    & $C_{FPMB}$ $(10^{-1})$ &$9.90\pm0.01$     \\
& $C_{NICER}$ $(10^{-1})$ &$7.95^{+0.04}_{-0.06}$    \\
& $\Delta\Gamma$ $(10^{-2})$ &$-9.6^{+0.9}_{-0.7}$  \\
\textsc{edge}       &E ($10^{-1}$ keV)    &$8.4\pm0.1$  \\
&$\tau_{max} (10^{-1})$            & $1.7\pm0.3$      \\
&E ($10^{-1}$ keV) & $4.5\pm0.1$\\
&$\tau_{max}$  &$1.2\pm0.4$ \\
\textsc{tbfeo}       & $N_{H}(10^{21}\ \rm cm^{-2})$ &$6.7\pm0.4$          \\
&$A_{O}$          &$1.3\pm0.1$         \\
&$A_{Fe}$         &$1.6\pm0.2$          \\
\textsc{diskbb} & $kT$ (keV)          & $1.29\pm0.01$  \\
&norm       &$77\pm3$       \\
\textsc{powerlaw} & $\Gamma$         & $3.05^{+0.03}_{-0.02}$   \\
&norm            & $1.2\pm0.1$        \\
\textsc{relxillns} & $q$   &$2.37^{+0.04}_{-0.05}$       \\
&$i$ ($^{\circ}$) & $3.1^{+5.2}_{-0.1}$          \\
&$R_{in}$ (\risco)        & $1.1\pm0.1$         \\
&$R_{in}$ (\rg) & $6.6\pm0.6$                  \\
&$kT_{bb}$ (keV) &$2.15\pm0.01$ \\
&log $\xi$       &$3.02^{+0.05}_{-0.04}$ \\
&$A_{Fe}$               & $4.7^{+0.5}_{-0.8}$  \\
&$\log{n_{e}} (\rm cm^{-3})$    &$18.98^{+0.02}_{-0.18}$   \\
&\textit{$f_{refl}$} & $0.22^{+0.03}_{-0.02}$    \\
&norm$(10^{-3})$     & $2.57^{+0.04}_{-0.06}$                      \\

\hline
$\chi^2$ (dof)             & &500.33 (390)

\end{tabular}

\medskip 
\end{center}
\end{table}

\subsubsection{Reflection Modeling} \label{ssec:reflection}
We used \textsc{relxillns} to model the reflection components of the combined spectra. The model as described in \citet{Garc_a_2022} is a fully self-consistent reflection model created specifically for characterizing the reprocessed emission from accretion disks of neutron stars. The model assumes illumination of the disk by a blackbody spectrum of temperature $kT_{bb}$. The inner and outer emissivity indices were tied together to create a single emissivity index ($q$). The redshift and dimensionless spin parameter were fixed at 0, since the source is galactic and the spin frequency of the source is not known. The inner disk radius, \rin, is allowed to vary and is returned in units of \risco, while the outer disk radius is fixed at the maximum value of 1000 \rg. The reflection fraction, $f_{refl}$, is set to positive values which means that the model contains both the illuminating blackbody in addition to the reprocessed spectrum. Thus, we remove the single-temperature blackbody component from the continuum model so that it is not model twice within the overall description of the spectra. The inclination ($i$), ionization parameter ($\log \xi$), iron abundance ($A_{Fe}$), electron density of the disk ($\log n_{e}$), and normalization are all free to vary.

Table \ref{tab:model_fit} shows the parameter values when the reflection model is applied to the spectra. Our model fit gives a reduced $\chi^2$ = 1.25.
We find a primary blackbody temperature of $2.15\pm0.01$ keV. This is significantly higher than the previously found value in \citet{Ludlam_2018}, where a temperature of $\sim$ 1.8 keV was reported. We find an emissivity index of $q = 2.37\pm0.03$, which agrees well with \citet{Miller_2013,Ludlam_2018,chiang2016evolution}.  
Our fit gives a disk inclination of  $i\le 8.3^{\circ}$ which fits well within the range of previously reported values reported from reflection studies as stated in \S \ref{sec:intro}. We find that the accretion disk extends to the NS surface as the inner disk radius is almost at  1 $R_{\mathrm{ISCO}}$. We note a supersolar abundance of iron in the fit. 
This phenomena has been well documented and is known to be degenerate with the electron density of the disk (see \citealt{Tomsick_2018,Garcia_2018_Fe_Abund} for more detailed discussions). 
\textsc{relxillns} has a hard limit for disk density of ${\log n_e} = 19$, but calculations show that disk density of accretion disks around compact objects should be on the order $\sim 22-23$ \citep{Garcia_2018_Fe_Abund}. We find a disk density consistent with the upper limit in \relxillns ($\log n_e = 18.98^{+0.02}_{-0.18}$), thus the enhanced Fe abundance is not unreasonable. The disk is moderately ionized at $\log \xi = 3.02_{-0.05}^{+0.04}$. Figure \ref{refl_allcomp}(a) shows the spectra with model components and ratio of the overall model to the data.

\subsection{Inspection of \nicer spectra}

\subsubsection{Investigation of Edges}
In this section we look at the effect that including {\sc edge} components may have on potential feature detection around the low energy feature. Due to the proximity of the edge components to our region of interest around 1.1 keV, we examined the impact that excluding them would have on our model fits (continuum and reflection). Figure \ref{Edge Comparison} shows the comparison of the best fits using the double thermal continuum model (Model 1) both with and without \textsc{edge} components. As can be seen, the model fit above $\sim$ 5 keV is in agreement regardless of whether or not edges are used in the fit. The edges are very clearly visible in panel (b). Their energies correspond with known detector features, namely the O K (0.56 keV), Fe L (0.71 keV), and Ne K (0.87 keV) edges.\footnote{\href{https://heasarc.gsfc.nasa.gov/docs/nicer/analysis_threads/arf-rmf/}{\nicer Responses}} As can be seen in Figure \ref{Edge Comparison}(b), removing the edge components did not result in any change in prominence of the potential 1.1 keV feature. When not including the \textsc{edge} components, the largest change of parameter values came from our multiplicative component \textsc{tbfeo}, with the best fit giving a higher hydrogen column density $(N_H)$ of $8.7 \times 10^{21}$ cm$^{-2}$ with a corresponding drop to the absorption abundances of O and Fe to 1.1 and 2.3, respectively. The resulting fits of the additive components were all in agreement within errors to the values found with edges present. Given the lack of significant change in our parameter values, we proceed by including the edges during our subsequent fits and analysis. We note that the presence of edges has been utilized in previous literature (\citealt{Ludlam20, Ludlam_2021, Moutard_2023,2024_Moutard_b}: for an in-depth discussion of these features see \citealt{2024_Moutard_b}).

\subsubsection{Archival Data}
Since the previous results in \cite{Ludlam_2018} reported the detection of the 1 keV feature, we analyzed the archival \nicer data from the 2017 observations. The 2017 \nicer data set (ObsIDs 1050320101-113) was reduced using the same {\sc nicerl2} calibration and filtering method, as well as spectral extraction as described in \S 2.1. The data were merged into a single spectrum as per \cite{Ludlam_2018}, but there is no clear evidence of the Fe L emission feature in the spectrum when applying the simple continuum model of Model 1. We further separated the dataset by intensity to investigate whether the feature was intensity dependent. Observations were grouped into intensity ranges of low (200-600 cts/s), medium (800-1200 cts/s), and high ( $>$ 1200 cts/s). 

\begin{figure}[t]
    \centering
    \includegraphics[width=\linewidth]{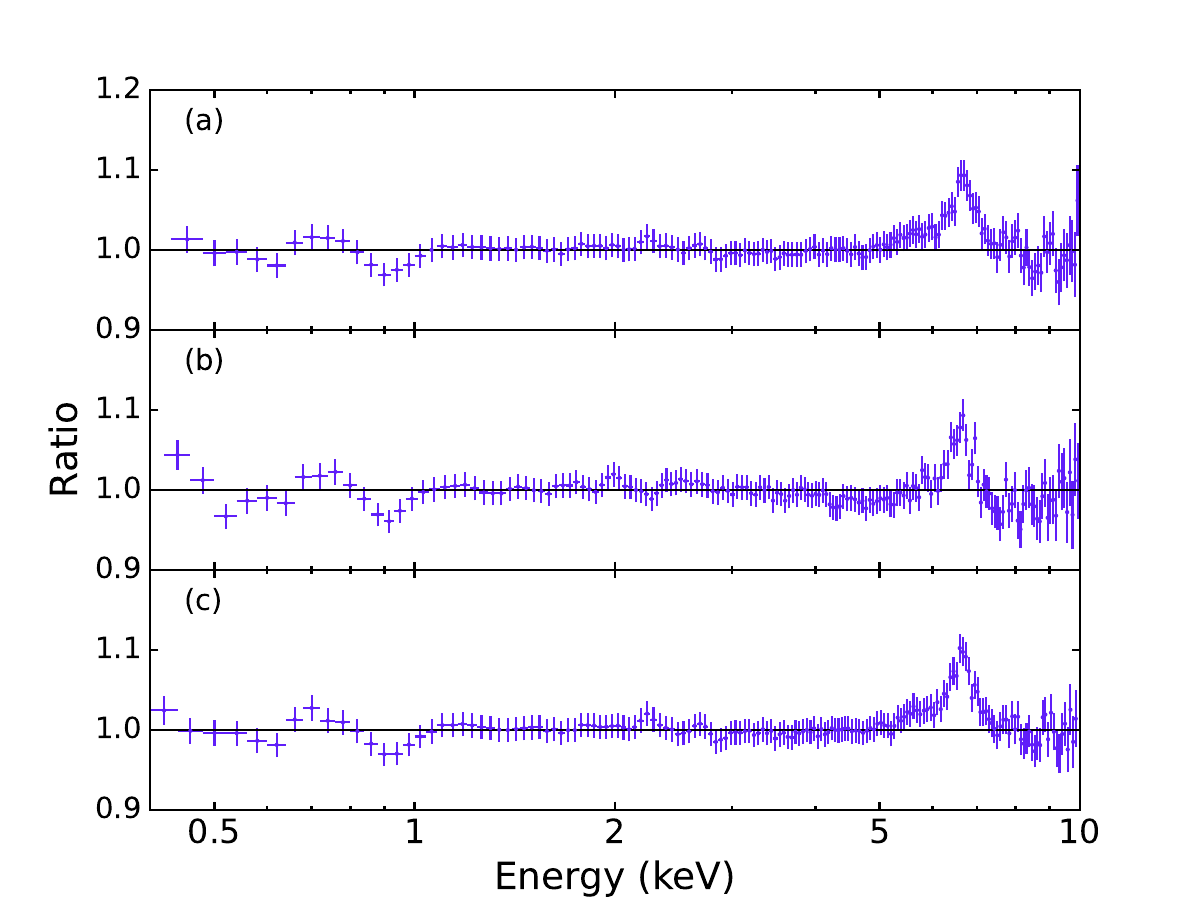}
    \caption{Ratio comparison of merged 2017 NICER Observations. The top panel (a) shows the total combined 2017 data. The middle panel (b) shows the high luminosity (1200-1700 cts). The bottom panel (c) is a mix of low (200-600 cts) and medium (800-1200 cts) luminosities. The continua were all fit with the {\sc xspec} model $\textsc{edge}\times\textsc{edge}\times\textsc{tbfeo}\times(\textsc{bbody}+\textsc{diskbb}+\textsc{powerlaw})$.}
    \label{archival data}
\end{figure}

Figure \ref{archival data} shows the ratio plots of the data to a simple continuum model in different intensity ranges. Again, no clear evidence is seen for the existence of the 1 keV feature in any of the intensity ranges. As a result, no claims can be made as to the intensity dependence of the emission feature. We do see a prominent Fe K$\alpha$ centered around 6.7 keV in each panel. We remark on the lack of the Fe L feature further in the following discussion section.
The continuum fits of the archival data give an inferred blackbody temperature of $kT_{bb} = 1.88\pm0.02$ keV with normalization of $4.46^{+0.02}_{-0.03}$. Using a disk component temperature of 1.15 keV, we find a normalization of $102^{+7}_{-5}$ $\rm km^2/$(D/10 kpc)$^2 \cos(i)$. Our fits give a power law photon index of $\Gamma = 2.8\pm0.3$ with normalization $0.59\pm0.1$ photons keV$^{-1}$. These values are consistent with those reported in \citet{Ludlam_2018}.

\subsection{XMM-Newton/RGS}
In addition to investigating the archival \nicer data, we analyze the XMM-Newton/RGS spectra. Due to the high energy resolution of the RGS, its passband coverage in the low energy range, and its peak effective area at 1.1 keV, we test for the existence of a relativistically broadened Fe L feature jointly with \nicer and \nustar. We modeled the continuum emission for all three RGS observations using both Model 1 and 2 from above to investigate the presence of the previously reported low energy feature $\sim$ 1.1 keV from \citet{Ludlam_2018}. In order to obtain a higher signal-to-noise ratio, we combine all three observations using {\sc rgscombine} similar to ISM studies performed in \citet{2013_Pinto} and \citet{2023_Psaradaki}. A feature near 1 keV was previously reported with XMM-Newton data with the EPIC PN camera \citep{2010_Cackett_et_al}, however it is subject to pile-up effects which could potentially artificially inflate the line profile and we discuss these implication in \S \ref{sec:discussion}. 
Due to the limited passband of the RGS (0.5--2.5 keV), when modeled alone with Model 1 and 2, several parameter values are poorly constrained and result in highly unphysical values due to the degeneracy of the different model components in the narrow energy range available. Applying a single continuum component (e.g., blackbody or power law) to search for the presence of an emission line, results in a roughly 4\% feature around 1.1 keV only when a power law was used. When modeled with a Gaussian, we find that the feature has a centroid energy of $1.10\pm0.02$ keV and $\sigma = 0.16^{+0.01}_{-0.02}$ and equivalent width of $6.4\pm1.0$ eV. When modeled by a blackbody, no presence of a line or emission feature was seen.

\begin{table*}[!t]
\begin{center}
\caption{Continuum Model Fits using RGS data}
\label{tab:cont_model_fit_RGS}
\begin{tabular}{llcc}

\hline
& & \multicolumn{2}{c}{NICER+NuSTAR+RGS}\\
\textbf{Component}                  & Parameter         & Model 1 & Model 2\\
\hline
\textsc{crabcor}    & $C_{FPMB}$ $(10^{-1})$    &$9.89\pm0.01$ & $9.90\pm0.01$ \\
& $C_{NICER}$ $(10^{-1})$ & $8.6\pm0.01$ & $8.6\pm0.1$  \\
& $C_{RGS}$ $(10^{-1})$ & $7.8\pm0.01$& $7.8\pm0.1$\\
& $\Delta\Gamma$ $(10^{-2})$ & $-8.6\pm0.8$ & $-6.3\pm1.0$\\
&$\Delta\Gamma_{RGS}$ $(10^{-2})$ & $-9.1\pm1.0$ & $-8.6\pm0.8$\\
\textsc{tbfeo}       & $N_{H}(10^{21}\ \rm cm^{-2})$ & $7.2^{+0.2}_{-0.1}$ & $6.4^{+0.2}_{-0.1}$ \\
&$A_{O}$          & $1.1\pm0.1$& $1.2\pm0.1$ \\
&$A_{Fe}$         & $1.1\pm0.2$& $1.2\pm0.4$\\
\textsc{diskbb} & $kT$ (keV) & $1.61\pm0.02$ & $1.98\pm0.03$ \\
&norm       & $36\pm 1$ &  $17\pm1$\\
\textsc{bbody} &$kT$ (keV) &$2.33^{+0.02}_{-0.03}$ & ...  \\
&norm$(10^{-2})$    & $1.9\pm0.1$ & ... \\
\textsc{powerlaw} & $\Gamma$  &  $3.34^{+0.04}_{-0.05}$  & ...\\
&norm     &  $1.1\pm0.1$  &  ...    \\
\textsc{nthcomp} & $\Gamma$ & ...& $1.88\pm0.02$  \\
&$kT_e$ (keV) & ...& $2.8\pm0.1$  \\
& $kT_{bb}$ (keV) & ...& $0.01\pm0.01$  \\
& norm  &... &$0.9\pm0.1$  \\
\hline
$\chi^2$ (dof)             & & 6783.95 (2232) & 6196.52 (2232)

\end{tabular}

\medskip 
\end{center}
\end{table*}

We then fit the data in conjunction with \nicer and \nustar using Model 1 and Model 2. We do not expect the values of galactic absorption to vary across observations or between instruments and as a result we tie these values when modeling the data together. We note that the fitting results with both current and 2018 calibration are identical both in terms of parameter values and fit residuals. Table \ref{tab:cont_model_fit_RGS} shows the result of the continuum fits using Model 1 and Model 2 with the RGS data.

We want to make particular note that with the addition of the RGS, the absorption abundances for oxygen and iron agree with values found in \citet{Ludlam_2018,2013_Pinto,2016_Gatuzz,2023_Psaradaki}. In particular, the iron abundance drops by 50$\%$. Figure \ref{2023 Cont}(b) shows the ratio of the RGS and \nicer to continuum Model 1 from 0.5 -- 3.5 keV.

We also performed a reflection fit to investigate the impact that the addition of the RGS data set would have on best fit parameters, if any. Similar to above, we model the combined spectra of all three instruments to obtain a reflection fit that simultaneously describes the complete spectrum from 0.4--30 keV which can be seen in Figure~\ref{refl_allcomp}(b). We use the same reflection model as \S \ref{ssec:reflection} and find that many of the parameters remain unchanged. However, some notable changes include the iron absorptive abundance in {\sc tbfeo} drops to 0.7, a parameter we believe could potentially be affecting the structure of low energy features, which we discuss further in \S \ref{sec:discussion}. Our best fit inclination remains low ($\sim 15^{\circ}$) and the disk iron abundance (A$_{Fe}$) drops to 2, which is closer to the expected solar value. The full results can be seen in Table~\ref{tab:model_fit_RGS}. 

Although the overall fit is formally unacceptable driven by the residuals in the RGS data (likely due to ISM features not being properly modeled in the overall fit), importantly, the key parameter of our inferred inner disk radius remained the same with the addition of RGS. 
Even when adding additional edge components to the RGS to better model the O and Fe L edges (see figure~3 in \citealt{2013_Pinto}), the results do not change.
We examined the model reflected spectrum from \relxillns, and similar to the \nicer and \nustar fit in Figure \ref{refl_model_only}, a noticeable broadened feature was seen in the 1.1 keV region. Assuming that the Fe L line is likely to be relativistically broadened similar to the Fe K, we also investigated whether the reflection model would produce similar parameter values if the Fe K region was ignored. However, the best fit values for many of the key parameters without the Fe K region are highly unphysical. We conclude that even when the reflection model does predict lower energy features, they are not strong enough on their own to accurately parameterize the spectrum and produce physically meaningful results.
\\

\begin{table}[!t]
\begin{center}
\caption{Reflection Model Fit of NICER-NuSTAR-RGS }
\label{tab:model_fit_RGS}
\begin{tabular}{llc}

\hline
Model                   & Parameter         & \\
\hline
\textsc{crabcor}    & $C_{FPMB}$ $(10^{-1})$ &$9.90\pm0.01$     \\
& $C_{NICER}$ $(10^{-1})$ &$8.07^{+0.13}_{-0.06}$    \\
& $C_{RGS}$ $(10^{-1})$ &$7.34^{+0.12}_{-0.06}$    \\
& $\Delta\Gamma_{NICER}$ $(10^{-2})$ &$-9.8^{+0.9}_{-0.6}$  \\
& $\Delta\Gamma_{RGS}$ $(10^{-2})$ &$-9.6^{+0.9}_{-0.7}$  \\
\textsc{tbfeo}       & $N_{H}(10^{21}\ \rm cm^{-2})$ &$6.9\pm0.1$          \\
&$A_{O}$          &$1.2\pm0.1$         \\
&$A_{Fe}$         &$0.7\pm0.1$          \\
\textsc{diskbb} & $kT$ (keV)          & $1.20\pm0.01$  \\
&norm       &$108\pm5$       \\
\textsc{powerlaw} & $\Gamma$         & $2.81\pm0.03$   \\
&norm            & $0.5\pm0.1$        \\
\textsc{relxillns} & $q$   &$2.87^{+0.12}_{-0.26}$       \\
&$i$ ($^{\circ}$) & $15.3^{+0.4}_{-3.5}$          \\
&$R_{in}$ (\risco)        & $1.1\pm0.1$         \\
&$R_{in}$ (\rg) & $6.6\pm0.6$                  \\
&$kT_{bb}$ (keV) &$2.15\pm0.01$ \\
&log $\xi$       &$3.02^{+0.05}_{-0.04}$ \\
&$A_{Fe}$               & $2.1^{+0.4}_{-0.3}$  \\
&$\log{n_{e}} (\rm cm^{-3})$    &$18.98^{+0.02}_{-0.18}$   \\
&\textit{$f_{refl}$} & $0.50^{+0.01}_{-0.10}$    \\
&norm$(10^{-3})$     & $2.36^{+0.08}_{-0.05}$                      \\

\hline
$\chi^2$ (dof)             & &4050.37 (2228)

\end{tabular}

\medskip 
\end{center}
\end{table}

\newpage

\section{Discussion} \label{sec:discussion}
We performed the first spectral analysis of simultaneous \nicer and \nustar observations of Serpens X-1 to characterize the full reprocessed emission from the accretion disk in the 0.4 -- 30 keV energy band. We tested the effect that various background estimation models would have on the spectra and found that, in this case, the use of the `3C50' and `SCORPEON' show no discernible difference in the appearance of reflection features. As such we can conclude that the spectral features are not background model dependent. We applied two different continuum model descriptions to the data and find that the presence of reflection features does not hinge of the choice of model. We applied the self-consistent reflection model \relxillns to the data to infer properties about the accretion geometry of the system. 

Our inferred inner disk radius of \rin = $1.1\pm0.1$ \risco ($6.6\pm0.6$ \rg for $a=0$) is consistent with the disk extending close to the NS as reported in other investigations \citep{Cackett_2008, 2010_Cackett_et_al, Miller_2013, Ludlam_2018,Matranga_2017,Bhattacharyya_2007,chiang2016evolution,Chiang_2016_FeK}.  
The truncation of the inner disk radius prior to \risco is thought to be driven to the existence of a boundary layer between the surface of the NS and the disk in Serpens X-1 \citep{chiang2016evolution}. It is also the origin of the single-temperature blackbody emission in the source spectrum. The radial extent of the boundary layer is directly proportional to the mass accretion rate ($\dot{M}$), which is proportional to luminosity \citep{Popham_2001}. Using our 0.4 -- 30 keV luminosity of 4.814$\times$10$^{37}$~erg s$^{-1}$, this corresponds to a mass accretion rate ($\dot{{M}}$) to be $1.9 \times 10^{-9}$ \(M_\odot\) yr$^{-1}$ using an efficiency ($\eta$) of 0.15 \citep{2002apa..book.....F}. Using Equation (25) in \citet{Popham_2001}, we found a maximum BL extent of 2.07 km or $\sim$1~\rg, which agrees well with our inferred inner disk radius, assuming a stellar radius R$_{*}$ = 10 km. 
We note that \citet{Mondal_2020} suggested that the disk could be truncated either by a boundary layer or a magnetic field. However, given the behavior of the inner disk radius and flux when investigating data over a large range in intensity with the same mission \citep{chiang2016evolution}, this strongly suggests that the disk is truncated by a boundary layer and not the stellar magnetic field.

The properties of the single-temperature blackbody component in our model fits of $kT\simeq2.46$ keV (continuum) and $kT\simeq2.15$ keV (reflection) agree well with values found in \citet{chiang2016evolution,Chiang_2016_FeK,Miller_2013,Mondal_2020}. We calculated the inferred emission radius from each of the thermal components in the continuum. For the single-temperature blackbody, assuming spherical emission, a distance of 7.7 kpc, and using a color-correction factor of 1.7 \citep{Shimura_1995}, an unphysically small emitting radius of 4.0 km was obtained from the blackbody normalization. If instead the emission arises from a banded equatorial region as expected for the boundary layer \citep{Inogamov_Sunyaev_1999}, the emitting radius ranges from R$_{\textsc{bb}}\sim$ 12.6 -- 17.8 km assuming the vertical height of the region is 5--10\% of the radius \citep{Ludlam_2021}. This is in agreement with the value obtained for the inner disk radius from the reflection model (\rin = $13.6\pm1.2$ km for 1.4 \ms). For the accretion disk component, we find an inner disk radius of $19.5\pm2.5$ km when using the {\sc diskbb} normalization. The {\sc diskbb} normalization has been shown to overestimate the inner disk radius when not accounting for the zero-torque boundary condition that is expected for thin disk accretion \citep{2005ApJ...618..832Z}. Additionally, using the disk temperature and normalization to infer inner disk radius has been shown to be an unreliable estimate in NS systems due to the spectral state degeneracy \citep{Degenaar_2018}.

The power law photon index of $\Gamma = 3.05^{+0.03}_{-0.02}$ is comparable to those found by \citet{Miller_2013,chiang2016evolution, Mondal_2020}, yet noticeably higher than that found in \citet{Ludlam_2018} (though the spectrum therein was normalized to the Crab and consequently could have a slope shift). Our best reflection fit preferred a lower inclination of $i \leq8.3^\circ$, which is comparable to \citet{Cornelisse_2013} who reported an inclination of $i \simeq 3^\circ$ based on optical observations. As mentioned in \S 1, the majority of investigations report inclinations ranging from $4.8^\circ \le i \le 30^\circ$ \citep{Cornelisse_2013, Miller_2013, Ludlam_2018,2010_Cackett_et_al,Cackett_2008,Matranga_2017}.
 
We find an emissivity index of $q = 2.37\pm0.03$, which is shallower than the value expected for flat, Euclidean geometry but agrees well with previous results \citep{Miller_2013,Ludlam_2018,chiang2016evolution}. Additionally, it is within the range of values seen for other accreting NS LMXBs (see \citealt{ludlam24} and references therein). We note that \citet{Wilkins_2018} found an index of $q = 3.5^{+0.3}_{-0.4}$ when looking at the 2013 \nustar data. It is important to highlight that that particular study centered only on the 3--10 keV, which biases the results around the Fe K$\alpha$ line rather than modeling the full reflection spectrum. 
Our ionization parameter is found to be log $\xi$ = $3.02^{+0.05}_{-0.04}$, which is less than the value reported in \citet{Ludlam_2018} of $3.2\pm0.1$ despite our higher inferred blackbody temperature of the NS/BL.

\begin{figure}[t]
    \centering
    \includegraphics[width=\linewidth]{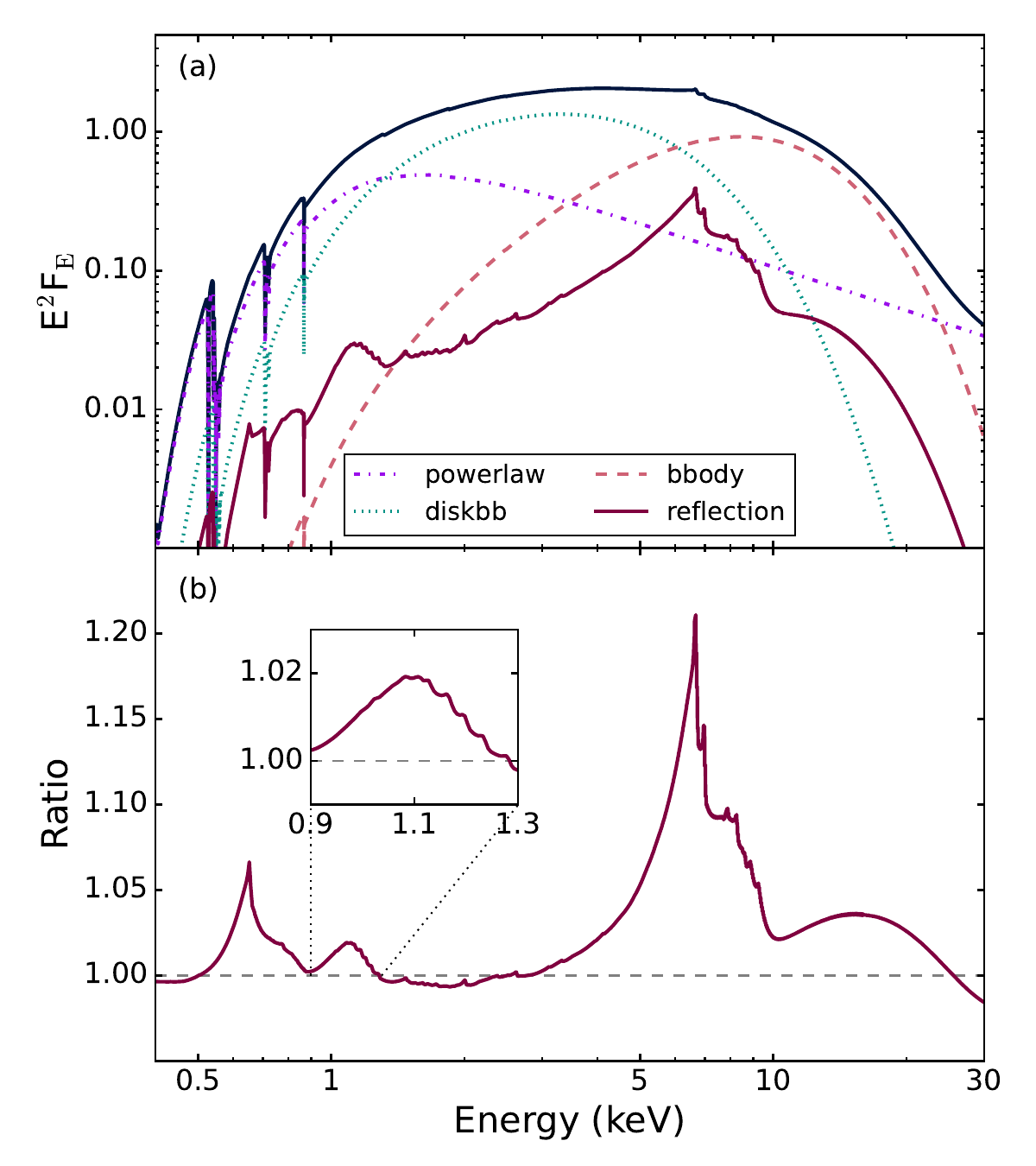}
    \caption{The unfolded model in the 0.4--30 keV energy range using the best fit parameters from Table \ref{tab:model_fit} is shown in panel (a). The overall model is shown as the black solid line. Note that the \relxillns model was switched to reflection only (i.e., negative $f_{refl}$) so it only contains the reflected emission, thus a blackbody was added for the input illumination. Panel (b) shows the ratio of the overall model to the individual continuum components so that the remaining contributions are solely from the reflection spectrum. There is evidence of the Fe L blend near 1.1 keV, but this is a $\sim2$\% feature relative to the reprocessed continuum. The feature at 0.68 keV is the oxygen Lyman$\alpha$ line. Both of these features are difficult to detect given the recommended systematic errors of 1.5\% on the \nicer data and potential instrumental artifacts at low-energy.} 
    \label{refl_model_only}
\end{figure}

The low energy feature is predicted by the model as can be seen in Figure \ref{refl_model_only} panel (a), however the feature is clearly not visible in the ratio plot of our continuum fits. Additional testing was done to explore what conditions may lead to an enhanced 1.1 keV feature in the spectrum. We found that fits favoring a lower disk temperature and subsolar absorption abundances of Fe in the {\sc tbfeo} model could produce an apparent low energy feature. When fixing the absorptive abundance of Fe to lower than typical values ($\le0.7$), a small, roughly 2\% feature is visible near 1.1 keV; yet this value is significantly lower than all best fit values found in \citet{Ludlam_2018} and the present paper. For Ser X-1, we know from the column density that the absorption primarily affects the lower energy region of the spectrum ($\lesssim$ 2.0 keV). Our primary fits with \nicer and \nustar gave absorptive abundances that were significantly larger than previously reported results. However, with the addition of the RGS, the values agreed with those found in \citet{Ludlam_2018,2013_Pinto,2016_Gatuzz,2023_Psaradaki}. Also, when modeling the archival \nicer data, we find values similar to those seen in the above literature. As shown previously, this can have an impact on residual structure in the resulting ratio plots. As a result, care should be taken when using this dataset and analyzing the reflection spectrum. Notably, the line profile is roughly $2\times$ weaker than the previously reported results of \cite{Ludlam_2018}.

One of the more impactful parameters in the \relxillns model was the reflection fraction, with higher reflection fractions strengthening all reflection features. Since the reflection fraction represents the ratio of intensity emitted towards the disk to that emitted toward infinity, we expect primary emission closer to the disk plane to result in higher fractions and in turn more pronounced reflection features. If the primary emission arises from the boundary layer and height is directly proportional to luminosity/mass accretion, then it is possible that during a less luminous observation the feature could be seen.  Additionally, Serpens X-1 would likely need to be in a harder spectral state (like the island state) with less thermal emission for any low-energy features to appear in the spectrum. Transient NS LMXBs may be ideal for searching for the 1 keV line as they can exhibit extreme hard states \citep{Parikh_2017} with multiple low-energy emission lines in their spectrum \citep{Ludlam_2016}.

After analyzing the joint \nicer and \nustar spectra and returning back to the archival data where the Fe L blend was previously reported, the lack of an emission feature at 1 keV could be due to a couple additional reasons: First, the initial analysis of these data in \cite{Ludlam_2018} was performed early in the \nicer mission when the calibration was still ongoing. Therefore, the data were normalized to residuals from the Crab Nebula to remove instrumental features from the spectrum since the Crab is a featureless spectrum. Residuals less than unity in the 1 keV range could inflate the Fe L line to be more prominent than it would be otherwise. As the mission calibration improved, the need to normalize to the Crab became obsolete. Second, the original analysis did not utilize any systematic errors on the dataset. Here, we use the systematic error values of in the CALDB for our spectra, which are at the 1.5\% level. 
Based on the feature's relative strength with respect to the reprocessed continuum (panel (b) of Figure \ref{refl_model_only}), when the \nicer systematic errors are used and absorptive abundances are allowed to fit freely, the detection of this feature may be extremely difficult, especially in softer thermally dominated spectra. However, this does not rule out the existence of said feature as it is clearly predicted by the model.  Since the presence of these detector features persists after all calibration and filtering files have been applied, compounded with the fact that the expected size of the Fe L line is near the systematics limit, it is unclear whether one would be able to confidently claim the existence of said line with present \nicer calibrations. 
Since we are unable to definitely claim detection of the feature in the spectrum, we cannot make any decisive claims about the dependence of the feature with source flux at this time.

Furthermore, we presented an analysis of the XMM-Newton RGS data as an additional test for the 1 keV line. In general, we found that when allowed to fit freely, the normalization of the disk blackbody component would be found at unphysically low or high values (depending on the model) compared to previous studies and our own findings with \nicer and \nustar. 
These fits would sometimes result in a feature appearing at roughly 1.1 keV in \nicer and RGS data sets. When applying a Gaussian to these specific fits, the centroid energies were found to be at a value of $1.10\pm0.02$ keV with width $\sigma = 0.16^{+0.01}_{-0.02}$ keV and normalization $1.4\pm0.2\times10^{-2}$ photons cm$^{-2}$. 
We find an equivalent width of $2.1\pm0.1$ eV. These are similar values to those found by \citet{1988_Vrtilek,Ludlam_2018, 2010_Cackett_et_al}.  
However, in fits where parameter values were more closely aligned with those in Table \ref{tab:cont_model_fit}, no clear visual feature was seen.
Despite the difficulties with the lower energy feature and a lack of clear detection, the addition of the RGS data set provided a further test and confirmation of the inner disk radius and inclination of Ser X-1.

The recent launch of the X-ray Imaging and Spectroscopy Mission (\emph{XRISM}; \citealt{XRISM_2018_Tashiro}) presents an exciting opportunity for high energy-resolution spectroscopy. The RESOLVE instrument has a spectral resolution of 5--7 eV and will be the ideal tool for investigating the accretion flow in NS LMXBs and the relativistic effects imparted on the reflection spectrum (Ludlam et al.\ 2025, in prep). Should the gate valve on RESOLVE successfully open, the energy resolution will be crucial for determining the presence of a 1 keV feature and the atomic species responsible. \\

{\it Acknowledgments:} 
This work is supported by NASA under grant No.\ 80NSSC23K1042. The material is based upon work supported by NASA under award number 80GSFC21M0002. This research has made use of the NuSTAR Data Analysis Software (NuSTARDAS) jointly developed by the ASI Space Science Data Center (SSDC, Italy) and the California Institute of Technology (Caltech, USA).

\bibliography{references}

\end{document}